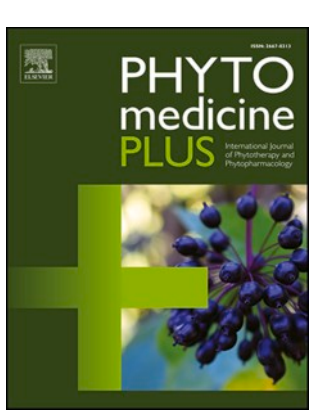

# Multi-ligand simultaneous docking of *Carica papaya* leaf phytochemicals, Carpaine and Rutin reveal multi-mechanism inhibition of cancer proteins, BCL-2 and WWP1

Merla Sudha [a], Asmita Saha [a], Belaguppa Manjunath Ashwin Desai [b], Anil Ranu Mhashal [c], Pronama Biswas [a,*]

[a] *Department of Biological Sciences, School of Basic and Applied Sciences, Dayananda Sagar University, Bengaluru, India*
[b] *School of Engineering, Dayananda Sagar University, Bengaluru, India*
[c] *Prescience Insilico Private Limited, Bangaluru, India*

ABSTRACT

Cancer remains a major global health concern due to chemotherapy resistance and toxicity from high-dose treatments. To overcome these challenges, new therapeutic strategies targeting key proteins in cancer progression are essential. This study evaluates two phytochemicals, Carpaine (Car) and Rutin (Rut), from *Carica papaya* leaves, for their potential in enhancing cancer therapy by targeting B-cell lymphoma 2 (BCL-2) and WW domain-containing protein 1 (WWP1) proteins. We assessed their additive, allosteric, and synergistic effects using molecular docking, multi-ligand simultaneous docking (MLSD), molecular dynamics (MD) simulations, and MMPBSA analysis. Car and Rut showed an additive effect on BCL-2 by binding at distinct regions within the same pocket. MLSD revealed an improved binding affinity of -13.13 ± 0.08 kcal/mol, individual ligands or the commercial inhibitor Venetoclax. For WWP1, Car bound near the H-site and Rut near the Le-site, exhibiting an allosteric effect that increased Car's binding affinity in MLSD to -15.59 ± 0.39 kcal/mol. Furthermore, Rut combined with bortezomib (Bortezomib) demonstrated a synergistic interaction with WWP1. Binding energies were -7.64 ± 0.156 kcal/mol for Bort, -10.26 ± 0.07 kcal/mol for Rut, and -15.59 ± 0.39 kcal/mol for MLSD, suggesting a more stable complex through synergy. These results suggest Car and Rut, particularly in combination with Bort, as promising candidates against cancer-related proteins BCL-2 and WWP1. Further experimental validation is warranted to explore their therapeutic potential.

## 1. Introduction

Cancer remains a major global health challenge, underscoring the urgent need for innovative treatment approaches. A key aspect of this involves understanding how anti-cancer drugs interact with their target proteins. Recent estimates suggest that breast, cervical, lung, and oral cancers are among the most common worldwide (Bray et al., 2024). The phenotypic heterogeneity of tumors allows some cancer cells to develop resistance to chemotherapy, often requiring higher drug doses, which can increase toxicity (Hamami et al., 2023). Targeting specific cancer proteins, such as the B-cell lymphoma 2 (BCL-2) (Kaloni et al., 2023; Zhang et al., 2020) and WW domain-containing E3 ubiquitin protein ligase 1 (WWP1) (Hu et al., 2021), which play crucial roles in apoptosis and cell cycle regulation, is a promising strategy.

BCL-2 is well known for its anti-apoptotic properties, enabling cancer cells to escape programmed cell death, which promotes tumor growth and metastasis (Kaloni et al., 2023). In breast cancer, elevated BCL-2 expression has been associated with chemotherapy resistance, highlighting its critical influence on cancer cell survival. Targeting BCL-2 has shown promise in improving therapeutic outcomes by restoring the ability of cancer cells to undergo apoptosis (Kønig et al., 2019; Merino et al., 2016). The BCL-2 family of proteins (consisting of anti-apoptotic members such as BCL-2 and BCL-xL, pro-apoptotic members like BAX and BAK and apoptotic regulators called BH3-only proteins) play a crucial role in regulating intrinsic apoptotic pathway. The anti-apoptotic proteins function by binding to the mitochondrial membrane and preventing the release of apoptogenic signals thereby ensuring cell survival. Increased level of BCL-2 is further associated with the reduction of pro-apoptotic proteins. This imbalance leads to resistance to apoptosis, thereby causing uncontrolled cell proliferation (Kaloni et al., 2023). BCL-2 is an important prognostic marker in ER-positive tumors, promoting tumor growth and metastasis in mammary gland development. Elevated level of BCL-2 is further observed in triple-negative breast cancer (TNBC) and luminal A subtype breast cancer (Merino et al., 2016). Venetoclax (Ven), the Food and Drug Administration (FDA) approved drug, had emerged as a promising treatment for malignancies characterized by elevated BCL-2 levels, including chronic lymphocytic leukemia (CLL), small lymphocytic lymphoma (SLL), and acute myeloid leukemia (AML). However, due to genetic instability, BCL-2 frequently undergoes mutations, leading to drug resistance and cancer relapses. Common mutations associated with Ven resistance include Gly101Val, Phe104Ile, Phe104Leu, Phe104Cys, Asp103Tyr, and Asp103Glu (Xu and Ye, 2022).

* Corresponding author.
*E-mail addresses:* pronama-sbas@dsu.edu.in, pronamabiswas@gmail.com (P. Biswas).

WWP1 has emerged as a key oncogene across various cancers, primarily by regulating protein degradation through the ubiquitin-proteasome system. It modulates the stability of proteins essential for cancer cell proliferation, migration, and metastasis (Lin et al., 2013). Studies indicate that silencing WWP1 in cancer cells results in growth arrest and increased apoptosis, emphasizing its role in maintaining cancer cell survival by suppressing apoptotic pathways (Lin et al., 2013). Ubiquitination, a crucial posttranscriptional modification, involves the attachment of ubiquitin molecules to proteins, marking them for degradation. This process relies on a cascade system comprising three key types of enzymes: ubiquitin-activating enzyme (E1), ubiquitin-conjugating enzyme (E2), and ubiquitin ligase (E3). WWP1 belongs to the NEDD4 family, a group of HECT-type E3 ligases responsible for transferring ubiquitin molecules. It is widely expressed across various tissues and plays a role in regulating essential biological processes such as receptor trafficking, protein degradation, damage response, and signal transduction (Hu et al., 2021). WWP1 has emerged as a significant factor in several cancers including breast, prostate, hepatocellular cancers and hematological malignancies, promoting cell growth, proliferation, autophagy and evasion of apoptosis. Overexpression of WWP1, degrades tumor suppressor genes, contributing to cancer progression (Lin et al., 2013). Bortezomib (Bort), a proteasome inhibitor, targets the synthesis of NEDD4 family proteins, including WWP1. However, its lack of specificity for WWP1 and limited efficacy have prompted the search for more targeted therapeutic approaches in cancer treatment (Behera and Reddy, 2023). Together, both BCL-2 and WWP1 represent crucial targets in cancer therapies aimed at reactivating apoptotic mechanisms and reducing tumor persistence.

Evaluating the effectiveness of cancer drugs requires understanding their interactions with target proteins (Fu et al., 2018). Conventional treatments like surgery and chemotherapy, though widely used, often weaken the immune system and increase susceptibility to secondary illnesses. Moreover, their high-cost limits accessibility for low-income individuals. These drawbacks have prompted the exploration of *in silico* approaches to identify affordable and effective phytochemical-based alternatives (Dibha et al., 2022). Setiawan et al. used *in silico* techniques to study the anticancer potential of compounds such as quercetin, scopoletin, and eleutherol from *Kleinhovia hospita Linn.* against breast cancer proteins (Kellik Setiawan et al., 2025). Similarly, Alifiansyah et al. investigated alizarin from *Rubia tinctorum* using QSAR modelling, molecular docking, and ADMET profiling to assess its activity against the MMP-9 receptor (Alifiansyah et al., 2024).

Computational techniques like molecular docking and molecular dynamics simulations are increasingly employed to predict protein-ligand interactions efficiently (Herdiansyah et al., 2024). Molecular docking predicts ligand binding affinities and modes of interaction with target proteins by simulating 3D models of protein-ligand complexes. This technique identifies active binding sites and evaluates ligand conformations through energy scoring, aiding in the rational design of new drugs (Hakaman et al., 2025). By minimising system-free energy and ranking ligand poses, docking algorithms reveal optimal binding modes, making them valuable tools for preclinical drug discovery.

In this study, we employed MLSD, along with molecular dynamics (MD) simulations and Molecular Mechanics Poisson-Boltzmann Surface Area (MMPBSA) analysis, to investigate the interactions between Carpaine (Car) and Rutin (Rut), which are the phytochemicals from *Carica papaya* leaves, and the cancer proteins BCL-2 and WWP1. These phytochemicals have demonstrated anti-cancer properties in previous studies (Satari et al., 2021; Vien and Loc, 2017). Car is an alkaloid that exhibits toxicity with an $IC_{50}$ value ranging from 1.13 to 2.94 μg/ml in breast cancer cell MCF7, carcinoma cell KB, leukemia cell HL-60 cell and lung cancer cell LU-1 cell lines (Vien and Loc, 2017). Car triggers the intrinsic (mitochondrial) apoptosis pathway by downregulating anti-apoptotic BCL-2 expression and upregulating p53 and caspase genes. Papaya extracts rich in carpaine cause significant BCL-2 suppression (with one study noting ~63 % reduction) alongside increased cytochrome c release and caspase-3 activation (Mahrous and Noseer, 2023). Carpaine also suppressess NF-κB signaling by promoting proteasomal degradation of the NF-κB p65 subunit resulting in reduced BCL-2 expression (Zhang et al., 2025). Rut is a flavonoid that inhibits the proliferation of lung, breast, prostate, and colon cancer by regulating several signaling pathways such as MAPK, TGF-β2/Smad2/3Akt/PTEN, Ras/Raf, and PI3K/Akt, which are associated with the induction of apoptosis and carcinogenesis. Rut has also been established as a drug that induces apoptosis and decreases drug resistance by exhibiting synergy when administered in combination with other drugs such as apigenin and tamoxifen (Satari et al., 2021). Rut promotes apoptosis by lowering BCL-2 levels and the BCL-2/Bax ratio, thereby tipping the balance toward cell death. For example, rutin treatment of cancer cells decreased BCL-2 while elevating Bax, which induces cytochrome *c* release and caspase activation (Perk et al., 2014). Rutin can also act through upstream regulators (e.g. p53 or tumor-suppressive microRNAs) that repress BCL-2 transcription (Huo et al., 2022). Notably, both compounds intersect with the WWP1 pathway: WWP1 is an oncogenic E3 ubiquitin ligase that normally helps tumors evade apoptosis (in part by sustaining BCL-2) (Behera and Reddy, 2023). Inhibiting WWP1 reduces BCL-2 and restores apoptosis, suggesting that carpaine's and rutin's pro-apoptotic effects may be further bolstered by targeting WWP1 in cancer therapy. Understanding such multitargeted actions not only validates these compounds as potential adjuvant therapies but also sheds light on strategic "Achilles' heels" of cancer (like BCL-2 dependence and WWP1-driven PTEN inactivation) that can be exploited for treatment. By harnessing compounds like carpaine and rutin, it may be possible to synergistically block cancer's survival pathways and trigger latent apoptotic processes, offering a natural, multi-pronged approach to cancer therapy. By employing MLSD, we aimed to explore the combined inhibitory mechanisms of these compounds on cancer proteins, providing insights that could guide the development of new therapeutic strategies.

## 2. Methods

### *2.1. Protein structure acquisition and validation*

The three-dimensional crystal structures for proteins, BCL-2 (PDB code: 4IEH) and WWP1 (PDB code: 6J1X) were obtained from RCSB PDB with resolutions of 2.1 Å and 2.3 Å, respectively. The quality of the proteins was assessed using PROCHECK (UCLA, Los Angeles, CA, United States) and VoroMQA (Vilnius, Lithuania) websites. In PROCHECK, Verify3D and Ramachandran plot tools were used. The selected structures were subjected to stringent criteria, ensuring that a Ramachandran score surpassed 90 %, the Verify 3D value exceeded 80 % (Sahu et al., 2013) and the Voromqa Global plot position remained above the worst 5 % line (Saha et al., 2024).

### 2.2. Homology modelling to fix missing residues

The missing residues were fixed using MODELLER. The full-length protein sequence was obtained in FASTA format from RCSB PDB. A Python script from Modeller was used to extract the protein sequence from the PDB file to create a .seq file. The sequence extracted from the PDB file was compared to the full-length sequence using pairwise alignment via the EMBOSS Needle tool, using EBLOSUM62 matrix, Gap penalty of 10 and extended penalty of 0.5 reveal the missing residues, facilitating the creation of an alignment file in .ali format required for residue modelling. Another python script was then used using Modeller to fill the missing residues using the alignment file. The best from all generated models was chosen based on the best DOPE (Discrete Optimized Protein Energy) score, which evaluates the overall quality of the protein model. A lower DOPE score indicated a more favorable structure (Sahu et al., 2013). The repaired structure was then visualized in 3D and subsequently used for molecular docking.

### 2.3. Ligand structure acquisition

The three-dimensional molecular structures of Car and Rut were retrieved from PubChem in the Structure Data File (SDF) format (Biswas et al., 2025b), with compound IDs 442,630 and 5280,805, respectively. Fig. 1 illustrates the structures of both ligands, which were drawn using ChemSketch software. The structures of the commercial FDA-approved inhibitors for the selected proteins were retrieved from PubChem in SDF format, with compound IDs 169,441,756 ((R, R)-Hydroxy Des (boric acid) Bortezomib, an analog of bort, (Phase IV for WWP1) and 49, 846,579 (Ven, Phase IV for BCL-2).

### 2.4. ADMET analysis of ligands

Pharmacokinetic studies are essential in determining the potential of new molecules as effective drugs and evaluating any potential toxicity in the human body. Physicochemical properties, absorption properties like solubility, gastrointestinal (GI) absorption, and skin permeability, as well as drug likeness properties like Lipinski's rule, bioavailability, and synthetic accessibility, were determined using SwissADME. Additionally, Human intestinal absorption (HIA), Distribution, Excretion and Toxicity were assessed using ADMETlab 2.0. The Biopharmaceutical Classification System (BCS) categorizes drugs into four classes based on their water solubility and intestinal permeabilitv (Dahan et al., 2009). This is crucial for optimizing drugs into higher classes through methods like nanonization (Dahan et al., 2009; Khadka et al., 2014).

### 2.5. Molecular docking

The modelled proteins were prepared in AutoDockTools-1.5.7 by adding Kollman charges and polar hydrogen atoms and saved as a .pdbqt file. Similarly, 3D conformers of ligands in .sdf format were obtained from PubChem, which were then converted to .pdb format using Open Babel GUI version 2.4 (O'Boyle et al., 2011). The PDB file of ligands were then saved as a PDBQT files by adding Gasteiger charges and hydrogen atoms in AutoDockTools-1.5.7 (Forli et al., 2016; Morris et al., 2009). Molecular docking was performed using a script-based method from AutoDock Vina, using VS code in a conda environment. The commands used for single docking and MLSD have been mentioned in Table 1.

The docked output was later visualized in USCF Chimera for 3D and Discovery Studio 2021 Client for 2D representation (Biswas et al., 2025a). Single, sequential, and MLSD were performed to study protein-ligand interactions. In single docking, a single ligand was docked with the target protein to assess its inhibitory potential. The output of single ligand docking was used as the input file for sequential docking. Sequential docking involves docking one ligand with the protein followed by combining the docked complex into a single file using UCSF Chimera. The second ligand was then docked to the complex following the same procedure as single ligand docking. Sequential docking was performed in two combinations, which were Car followed by Rut, and vice versa. MLSD was used to analyze the binding of the two ligands docked with the protein at the same time. This method aimed to determine whether the ligands exhibited additive effects, synergistic behavior, or allosteric interactions with the protein. Multiple sets of dockings were performed to ensure consistency in the docking outcome. Fig. 2 schematically represents the methodology followed. The proteins were then docked with their commercially available inhibitors identified from ChemBL database to compare their binding affinities and interactions with Car and Rut. The docking methodology followed was consistent with the previously described protocol (Fig. 2). Notably, a bort analog was used in place of bort due to two primary reasons. Firstly, the 3D conformer of bort was not available in PubChem database and secondly, bort contains a boron atom which is not supported in the default force field of the docking software. This limitation arises from the absence of parameters required to model interactions involving boron rendering the software incapable of processing boron-containing ligands. Consequently, molecular docking was performed with a boron-free analog of bort. According to Vega-Valdez et al. (2021) bort and its analogs bind to the same active site and exhibit comparable inhibitory effects supporting the use of the analog for this study.

**Table 1**
Docking Commands for Single Ligand and Multiple Ligand Simultaneous Docking.

| Docking | Command |
|---|---|
| Single docking | vina_1.2.5_linux_x86_64 –receptor receptor.pdbqt –ligand ligand.pdbqt –config config.txt –exhaustiveness 32 –out output.pdbqt –num_modes 3 –verbosity 2 |
| Multi-ligand simultaneous docking | vina_1.2.5_linux_x86_64 –receptor receptor.pdbqt –ligand ligand1.pdbqt ligand2.pdbqt –config config.txt –exhaustiveness 32 –out output.pdbqt –num_modes 3 –verbosity 2 |

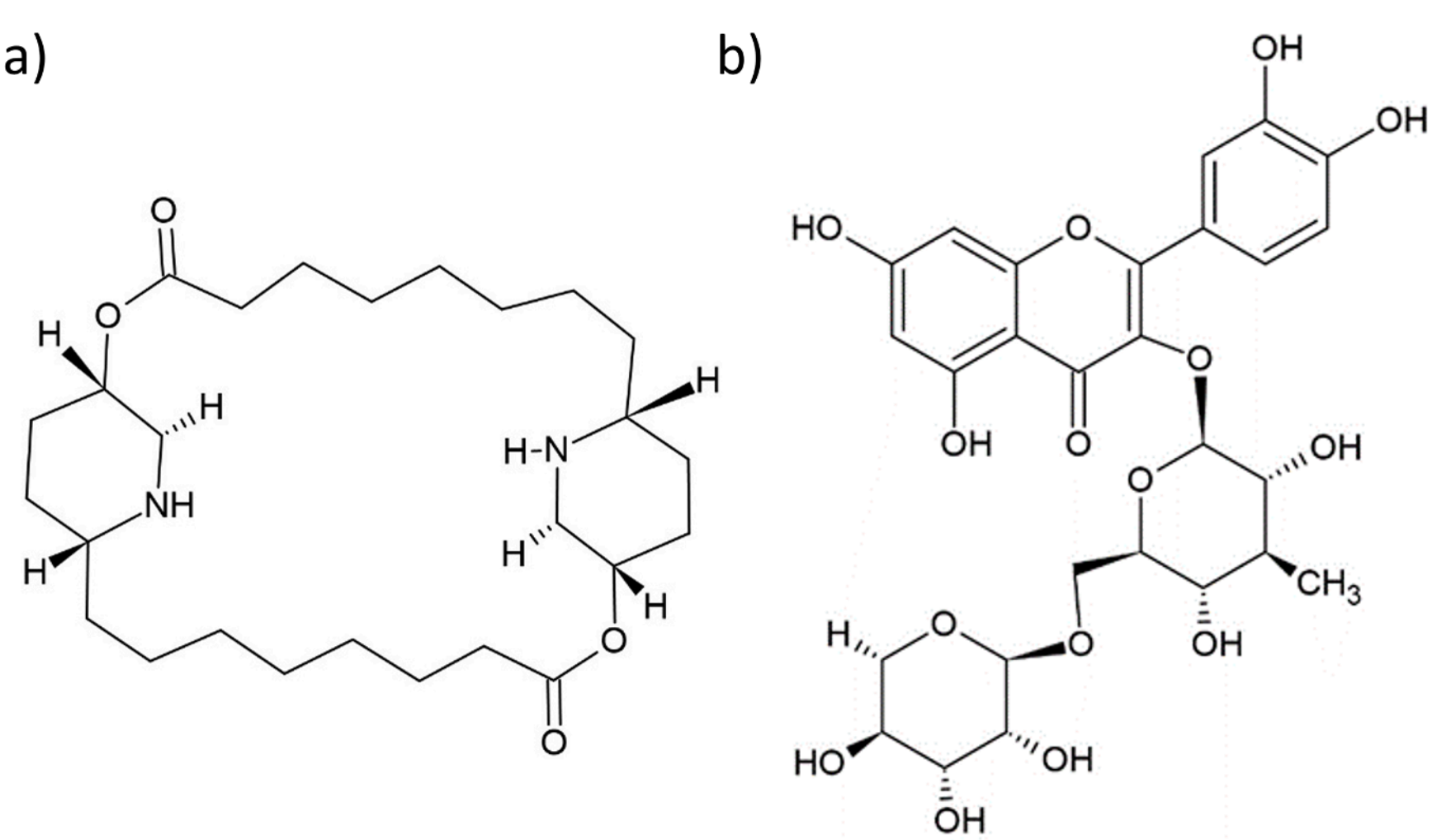

**Fig. 1.** Chemical structures of (a) Carpaine (Car) and (b) Rutin (Rut).

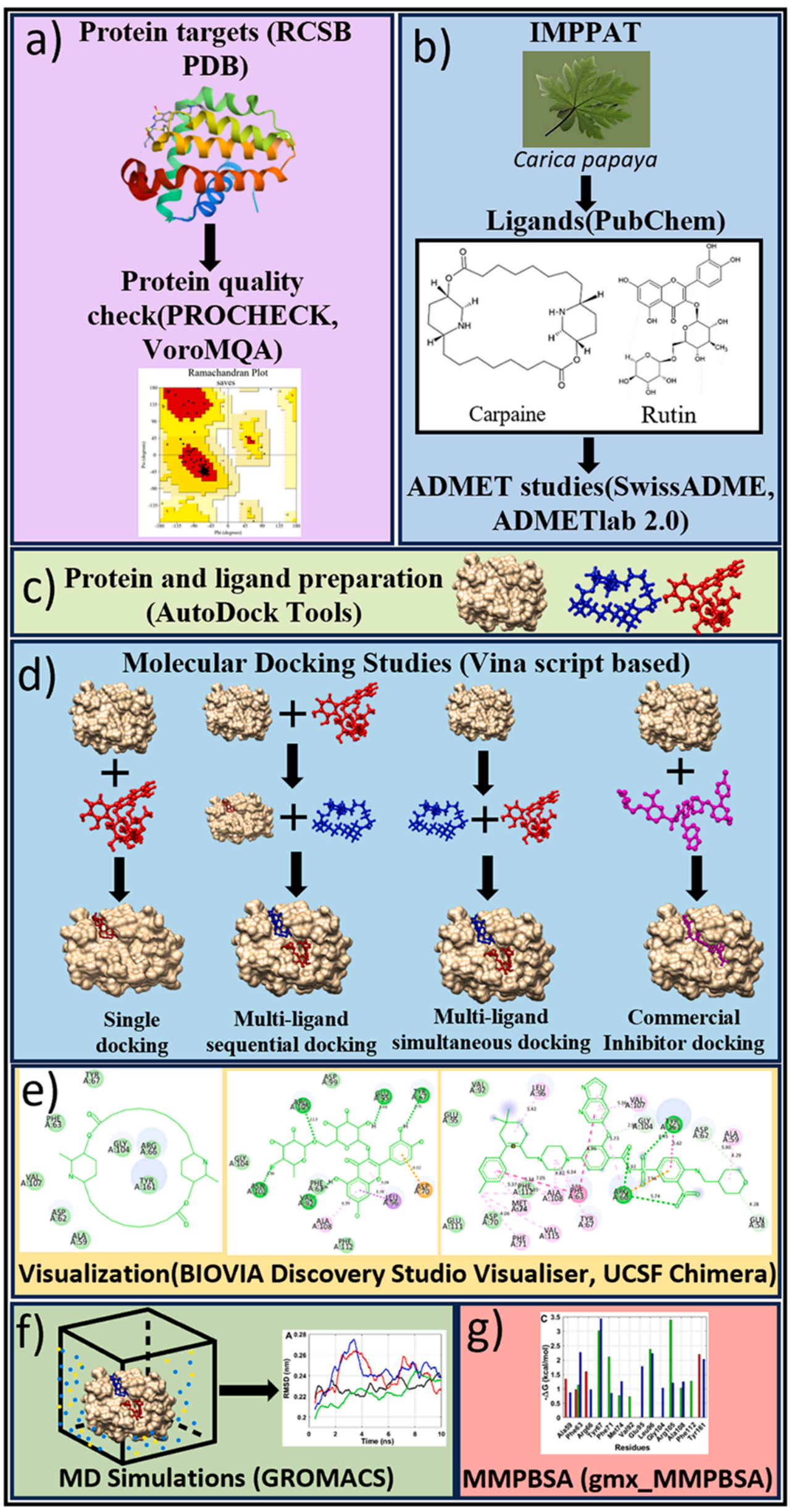

**Fig. 2.** Schematic flowchart for molecular docking and MD simulations. Proteins were sourced from the RCSB PDB and ligands from Indian Medicinal Plants Phytochemistry (IMPPAT) and PubChem. Protein structures were verified using Procheck and Voromqa, while ADMET studies were conducted on ligands. Structures were prepared with Auto-Dock Tools, and molecular docking was performed using Au-toDock in four modes: single, sequential, MLSD, and with inhibitors. This was followed by MD and MMPBSA analysis.

### 2.5.1. *Validation of docking protocol*

To validate the procedure and preparation of receptor and ligand molecules, as well as the docking protocol, RMSD (Root Mean Square Deviation) validation was conducted. This process involved redocking co-crystallized ligands of the selected proteins using the same methodology described previously, followed by comparing their RMSD with the native binding pose of the co-crystallized ligands in the protein .pdb file (Shahroz et al., 2022). UCSF Chimera was employed to superimpose the redocked ligands and the native poses of co-crystallized ligands and to compare their RMSD values. A lower RMSD value indicates a more accurate docking procedure. Typically, an RMSD value lower than 2 Å is acceptable. This rigorous validation process ensures the reliability and accuracy of the docking procedure, enhancing confidence in the predicted ligand-protein interactions (Morris and Lim-Wilby, 2008). The co-crystallized ligand of BCL-2, 1e9 (identified from RCSB PDB) was used for the RMSD validation. After redocking, RMSD result with the native and redocked pose was 1.692 Å. This can be considered as a good RMSD result thereby verifying our docking protocol.

## 2.6. *Molecular dynamic simulations*

MD simulations were conducted using GROMACS 2021.2 with the CHARMM27 force field. CGenFF was utilized to generate topology files for the ligands. The docked complexes were solvated in a dodecahedral box using the TIP3P water model, ensuring minimum distances of 1.0 nm and 3.0 nm from the box edges for BCL-2 and WWP1, respectively. Sodium and chloride ions were added to neutralize the system. Energy minimization was performed using the steepest descent algorithm with 50,000 steps. The system was equilibrated under both NVT and NPT ensembles, each for 100 ps (50,000 steps). The temperature was maintained at 300 K and the pressure was set to 1 bar. The MD simulations were performed for 100 ns, with a 2-fs time step. To ensure continuity and efficient data handling, the simulation was divided into five 20 ns segments using checkpoints, which were later merged. Structural stability and compactness were assessed using RMSD and radius of gyration (Rg), followed by an analysis of the MD simulation movies, which were generated using ChimeraX. Additionally, per-residue decomposition analysis was conducted on all frames of the MD trajectory using gmx MMPBSA to calculate free energy contributions (Valdés-Tresanco et al., 2021). MMPBSA estimates the total binding free energy of the ligand to the target protein by combining molecular mechanics with solvation terms derived from the Poisson-Boltzmann equation and surface area calculations. It also estimates the per-residue decomposition energy where the total binding energy is broken down into contributions from individual residues in the docked complex (Li et al., 2023). This analysis focused on residues within 4 Å of the ligand and employed the leaprc. protein. ff14SB force field for precise evaluation

# 3. Results and discussion

## 3.1. *Additive effect of car and rut observed in BCL-2*

Based on the number of ligands involved, docking can be classified into single-ligand, multi-ligand sequential, and multi-ligand simultaneous docking (MLSD). MLSD offers a more realistic approach by simultaneously considering multiple ligands, better reflecting the complex molecular interactions in biological systems (Li et al., 2014). In real protein-ligand binding processes in the host cell, multiple molecules, such as substrates, cofactors, ions, and water molecules influence each other. However, majority of the docking studies typically focus on one ligand at a time. MLSD offers a more realistic approach by enabling the simultaneous docking of multiple ligands, which better mimics molecular recognition processes and provides valuable insights into their combined action and potential synergistic effects, thereby enhancing our understanding of ligand-receptor interaction (Saha et al., 2024). Docking BCL-2 with Car and Rut revealed two distinct binding pockets as observed in Fig. 3a. Interestingly, both ligands also bound to the same pocket as Ven, as shown in Figs. 3c and 3d. In single docking, Rut demonstrated a higher binding affinity of −8.56 ± 0.005 kcal/mol compared to Car with −7.72 ± 0.002 kcal/mol (Table 2). Car formed van der Waals interactions with Gln 50, Ala 51, Asp 54, Phe 55, Arg 58, Gly 96, Val 99, and Tyr 153. Rut formed hydrogen bonds with Glu 65, Gln 69, Val 84, Glu 87, Arg 97, and Ala 100, hydrophobic bonds with Phe 55, Met 66, Leu 88, Arg 97, and Ala 100, and van der Waals interactions with Phe 63, Gly 96, Glu 103, and Phe 104, as shown in Fig. 3a. Sequential docking provided further insights into binding interactions. When Rut was docked after Car, the interacting residues of Car remained unchanged from single docking (Fig. 3b). However, when Car was docked after Rut, Rut formed additional interactions including a pi-sigma bond with Tyr 59 and alkyl bonds with Met 66, Leu 88, and Ala 100. These new interactions suggested that the presence of Car enhanced the binding of Rut by stabilizing the binding pocket and facilitated stronger interactions (Li et al., 2023). MLSD allows the simultaneous docking of multiple ligands into a protein-binding pocket. This method considers the cooperative and competitive interactions between ligands, providing a more comprehensive understanding of how multiple compounds can influence each other's binding (Saha et al., 2024). In MLSD, Car and Rut bound together with an increased affinity of −13.13±0.07 kcal/mol compared to the single docking with Ven, which had a binding affinity of −10.37±0.493 (Table 2). Both the ligands together occupied the same binding pocket as Ven, but they bound in two distinct fragments. This fragment-based binding allowed Car and Rut to better complement the binding site, filling it more completely and efficiently than Ven alone. Car and Rut's combination offers an improved fit with higher binding affinity, which covered all residue interactions observed with Ven. This can be further verified by comparing the interacting amino acids, highlighted with red circles in Fig. 3c and 3d Even though the binding pockets of Car and Rut were nearby, no inter-ligand interactions were observed in molecular docking and the ligands did not compete. The sum of interactions of both our ligands in MLSD was more effective than the commercial inhibitor and the single docking results of Car and Rut. From these observations, we predicted that Car and Rut showed an additive effect (Olszowy-Tomczyk, 2020).

Furthermore, a comparison of our results with previous molecular docking studies suggested that our combination of ligands exhibited

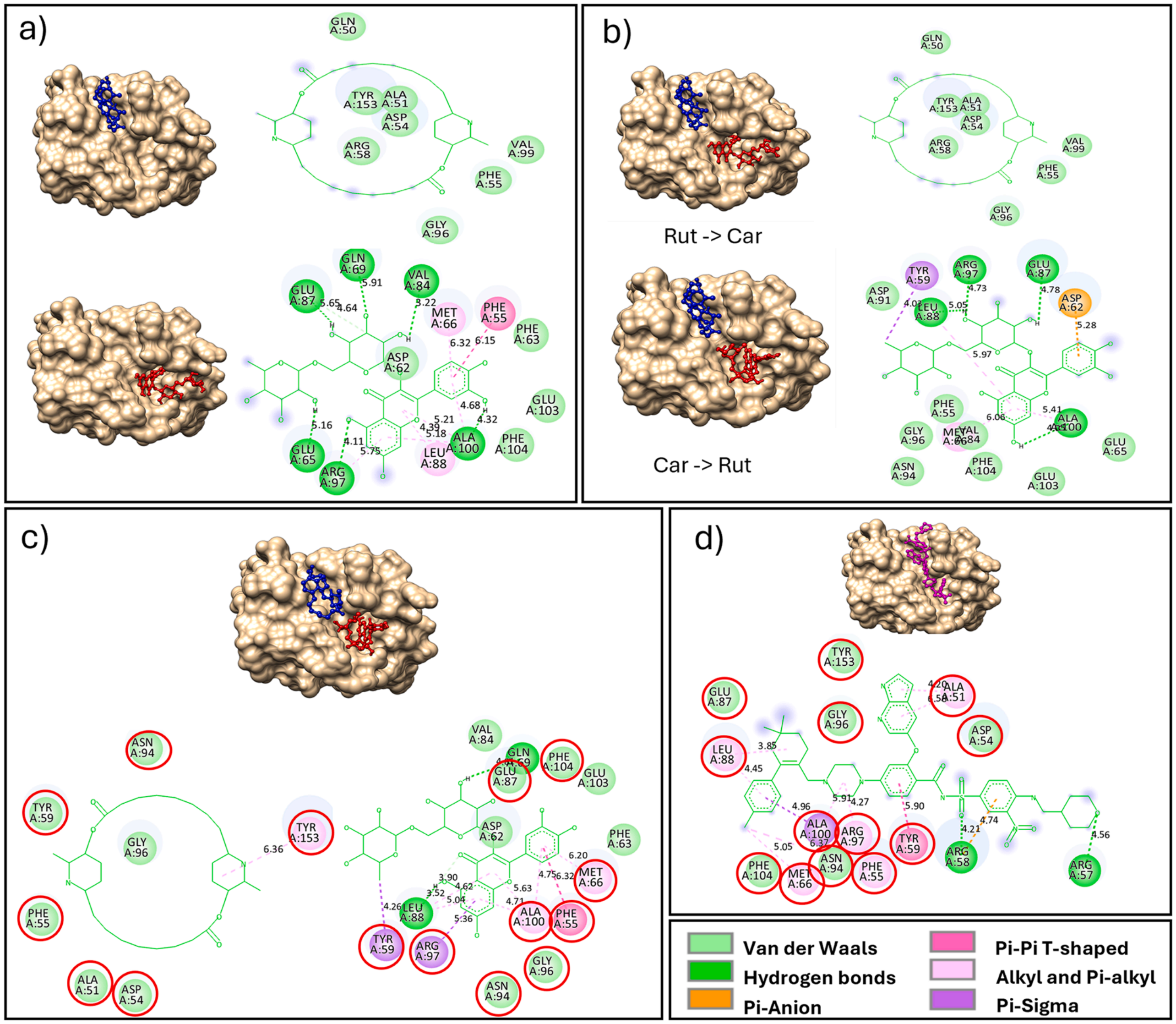

**Fig. 3.** Visualization of amino acid interactions and binding pocket of BCL-2. (a) Single ligand docking with Carpaine (Car)(blue) and Rutin (Rut) (Red). (b) Sequential ligand docking; Car and Rut bind in their respective pockets with slight confor-mational changes (c) Multiple ligand simultaneous docking (MLSD) (d) Docking with commercial inhibitor Venetoclax (Ven) (pink). The common amino acid residues between Car and Rut in MLSD and Ven are marked with red circles in the panels (c) and (d) Hence predicted additive effect is seen between Car and Rut.

**Table 2**
Predicted binding affinity values of selected ligands with BCL2 and WWP1 proteins.

| Protein | Single | | Sequential | | MLSD[1] | | Inhibitor |
|---|---|---|---|---|---|---|---|
| BCL2 | Car[2]<br>−7.72±0.002 | Rut [3]<br>−8.56±0.005 | Rut->Car[4]<br>−7.92±0.001 | Car->Rut[5]<br>−8.52±0.197 | −13.13±0.075 | | −10.37±0.493 |
| WWP1 | −9.12±0.01 | −10.26±007 | −9.22±0.01 | −9.97±0.019 | Car+Rut<br>−15.59±0.39 | Bort+Rut<br>−13.27±0.25 | −7.64±0.156 |

[1] Mutil-ligand Simultaneous docking
[2] Carpaine
[3] Rutin
[4] Rutin followed by Carpaine
[5] Carpaine followed by Rutn

comparable or even better BCL-2 inhibitory potential. For instance, Saha et al., conducted a study on *Moringa oleifera* leaf phytochemicals for their potential to inhibit BCL-2 using Multi-Ligand Simultaneous Docking (MLSD), to understand the anticancer properties of the leaf. MLSD analyses highlighted inter-ligand interactions among various bioactive compounds, such as Apigenin, Hesperetin, N,α-l-rhamnopyranosyl vincosamide, Niazimicin A, and Niaziminin B. The study explored several MLSD combinations with three and four phytochemicals docked simultaneously. Notably, the combination of Apigenin, Hesperetin, and Niazimicin A yielded the highest binding affinity of –14.96 kcal/mol (Saha et al., 2024). In comparison, our MLSD results employed only two ligands to obtain a binding affinity of –13.13 ± 0.075 kcal/mol, which is relatively close to the binding affinity achieved with the three-ligand combination mentioned above. Furthermore, the same study reported six other combinations with binding affinities lower than –13.13 kcal/mol. This underscores the strong inhibitory potential of Carpaine and Rutin, even when used in smaller combinations. Similarly, Biswas et al. performed MLSD of Garcinol and Withaferin A, which are phytoconstituents of *Garcinia indica* and *Withania somnifera* respectively with BCL-2. They reported a binding affinity of −11.88 ± 0.12, which is lower than our MLSD binding affinity value (Biswas et al., 2025b). Moreover, BCL-2 is prone to mutations due to genetic instability, often leading to drug resistance and cancer relapse. Notably, mutations such as Gly101Val and Asp103Tyr have been linked to resistance against Venetoclax (Saha et al., 2024). This underscores the urgent need for novel inhibitors capable of maintaining efficacy despite such mutations. Future studies should focus on this critical aspect by incorporating mutational analyses to evaluate the robustness of potential drug candidates

From the MD simulation analysis, it was seen that the RMSD of apo protein was high compared to the other docked complexes. This could indicate that the docked complexes are contributing to the stability of BCL2, further improving its inhibition. Moreover, we observed that the RMSD of single docked complexes with Car and Rut were lower compared to Ven, making it more stable. The Rg of the MLSD docked complex was observed to be lower compared to all the other docked complexes, further supporting the statement that the combination was better than the commercial inhibitor (Fig. 4a and 4b). Introducing a linker between the ligands could further improve RMSD and Rg. The MD simulation movie of Car [see Supplementary video file 1] revealed that it did not bind strongly at the active site, as it exited the pocket during the simulation. This weak binding can be attributed to the presence of only van der Waals interactions (Fig. 3a). Interestingly, MLSD remained stable with both Car and Rut bound at their respective sites [see Supplementary video file 2], further confirming the stability of the docked complex. Similarly, Rut and Ven were strongly bound within their pockets [see Supplementary video files 3 and 4]. The amino acids Phe 55, Met 66, Val 84, Glu 87, Leu 88, Val 99, Tyr 153 and Phe 104 exhibited higher per-residue binding free energy contributions in the MLSD condition compared to single docking (Fig. 4c). The binding free energy contributions of MLSD, car, rut, and Ven were −41.32 kcal/mol, −14.69 kcal/mol, −19.35 kcal/mol, and −32.52 kcal/mol, respectively. From these values, we inferred that the MLSD complex is more stable than the commercial inhibitor and the single dockings.

Previous studies have performed MD simulations on BCL-2; however, our study is the first to conduct MD simulations involving MLSD with two phytochemicals. For instance, Lv et al. investigated novel natural inhibitors targeting BCL-2. Their study showed that these inhibitors bind closely to BCL-2 and form stable complexes during MD simulations However, their MD was conducted for 30 ns and this simulation for such short duration may be insufficient to capture the complete dynamics, as key biological mechanisms take place on timescale of milliseconds or longer, necessitating extended MD simulations to observe the stability of the complex (Shaw et al., 2010). Additionally, Lv et al. used Obatoclax, a Phase 3 clinical trial compound, as the reference inhibitor. In contrast, our study employs Venetoclax, a Phase 4 (FDA-approved) inhibitor, thus aligning more closely with current therapeutic standards (Lv et al., 2021). Moreover, beyond evaluating inhibitory interactions, MD simulations have been applied in other studies under high-temperature unfolding conditions ranging from 400 K to 800 K for 25 ns. These simulations highlighted that the structural integrity and stability of the core of the BCL-2, and the flexibility of its Flexible Loop Domain (FLD), which are critical for its apoptotic regulatory function. Understanding these structural dynamics under stress conditions helps identify vulnerable regions within BCL-2 that may serve as strategic targets for drug design, especially for compounds aiming to modulate or disrupt protein–protein interactions involved in apoptotic signalling (Ilizaliturri-Flores et al., 2014).

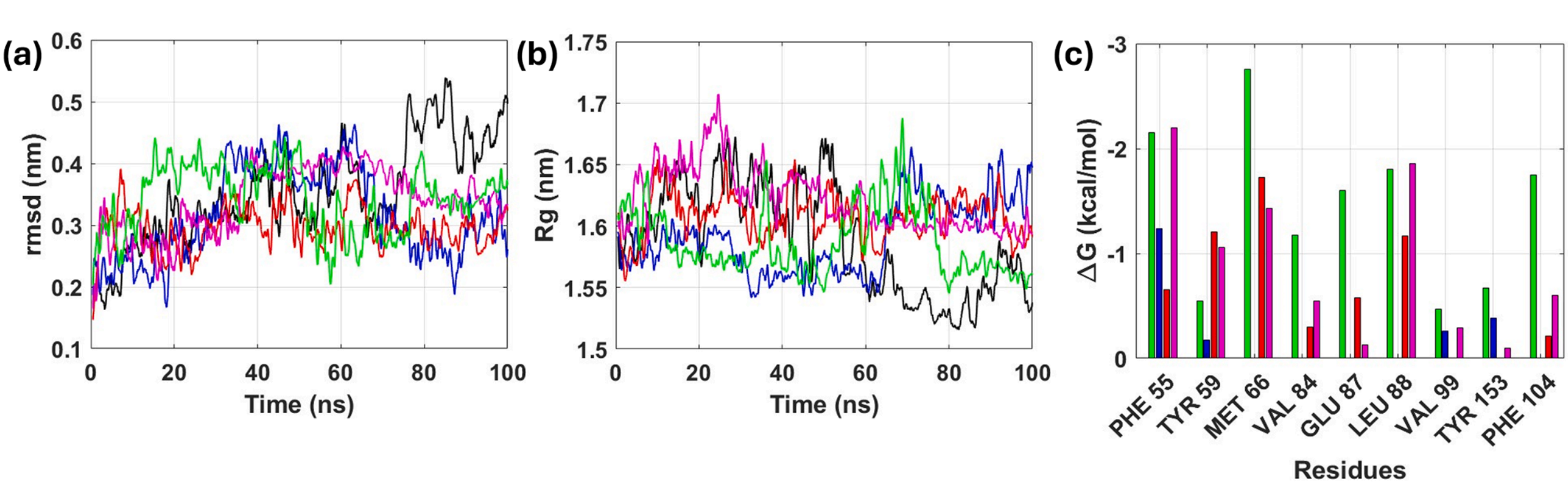

**Fig. 4.** (a) Root mean square deviation (RMSD) and (b) Radius of gyration (Rg) of all atoms computed for BCL-2. (c) Free energy decomposition of residues for single dockings and MLSD with BCL-2.

### 3.2. WWP1 is inhibited by allosteric interaction between car and rut

WWP1 possesses three regulatory regions known as the H-site (headband), *Le*-site (left ear) and *Re*-site (Right ear) (Fig. 5d) (Wang et al., 2019). In silico studies are being conducted to identify new inhibitors for this protein. Dudey et al. developed inhibitors for WWP1 and WWP2 using a structure-activity relationship (SAR) by synthesis approach, supported by molecular docking. Their study focused on designing inhibitors for these HECT E3 ubiquitin ligases, which are responsible for degrading tumor suppressor proteins and are often dysregulated in cancer (Dudey et al., 2024). Similarly, another study employed molecular docking to explore how Indole-3-carbinol (I3C) and its derivatives interact with WWP1 and WWP2. Docking results showed that 3,3′-diindolylmethane, which is a condensation product of I3C, bound at the ubiquitin exosite of WWP1 and forms hydrophobic interactions with the WW2 domain, contributing to its stronger inhibitory activity (Dudey et al., 2025). These findings highlight the importance of docking in understanding SAR and guiding the design of WWP1 inhibitors. Although such studies emphasize the inhibitory potential of WWP1, its activity is also influenced by allosteric regulation. Our study observed allosteric interactions with natural compounds Car and Rut. Instead of focusing solely on synthetic inhibitors or their derivatives, natural inhibitors may offer a promising alternative approach.

Our single docking analysis revealed that both Car and Rut bound near the H-site, which was a favorable pocket with common amino acids, including Arg855, Asn638, Met865, Tyr633, His621, Asp675, Leu856, Pro857, and Met865 (Fig. 5a). Rut exhibited a slightly higher binding affinity ($-10.26\pm007$ kcal/mol) compared to Car ($-9.22\pm0.01$ kcal/mol) (Table 2). The interacting residues and corresponding bonds for Car and Rut are shown in Fig. 5a. Sequential docking revealed a second binding pocket emerged near the Le-site, which we propose as the allosteric site (Fig. 5b). This intriguing observation raised the question of which ligand will occupy the favorable position when both ligands approach together. This was further investigated using MLSD, which revealed a higher binding affinity of $-15.59\pm0.399$ kcal/mol compared to all other docking scenarios. Notably, only Rut with higher affinity than Car bound near the preferred site H-site, while Car occupied the predicted allosteric site (Le-site). Furthermore, the analog of commercial inhibitor Bort bound near the H-site, sharing common amino acid interactions with Rut in MLSD, including Tyr 259, Thr 473, Cys 260, Gly 297, Thr 296, Arg 475, Thr 471, Pro 477, Asn 258, His 241, Asp 295 (Fig. 5c). These shared amino acids that are highlighted using red circles in Figs. 5c and 5d. MD analysis showed that the RMSD of apo protein was closer to MLSD which indicated that MLSD was as stable as the native structure of the protein (Fig. 6a). Moreover, the RMSD of MLSD was lower than that of the single docking of car which suggested that the docked complex was more stable and favorable compared to Car. Similarly, we observed that the Rg of the apo protein was higher compared to other docked complexes between 10 ns and 30 ns, but it gradually stabilized between 80 ns to 100 ns (Fig. 6b). Here, the Rg of the apo protein was slightly higher compared to the single dockings and MLSD. From this we inferred that after docking with our ligands, the protein gained slightly more stability compared to the apo protein. Additionally, we observed that the Rg of MLSD was higher than rut and similar to car which further supported that MLSD was more stable and favorable than single docking. From the MD simulation movies, we further confirmed the stability of the MLSD and single docking complexes, supporting the idea that Car and Rut can exhibit an allosteric effect to inhibit WWP1 [see Supplementary video files 5–7]. From MMPBSA, we observed that Phe 237, Leu 244, Tyr 259, Pro 477, Leu 478 and Met 485 were the common interacting amino acids among the single dockings of car, rut and MLSD (Fig. 6c). Leu 244, Tyr 259, and Pro 477 were found to have a higher free energy decomposition in MLSD compared to the single dockings. From this we inferred that since the amino acids in MLSD exhibited higher binding free energy the interaction was stronger which further contributed to its stability. The free energy of single docking of WWP1 with Car is $-18.24$ kcal/mol, with Rut was $-20.15$ kcal/mol and $-32.30$ kcal/mol in MLSD. From these findings, we inferred that Car and Rut exhibited allosteric interactions (Changeux, 2013). While they were primarily bound to the same pocket individually, a second regulatory pocket emerged as the binding site when a second ligand was introduced. Moreover, the binding pocket was the same in the case of single docking and with the analog of commercial inhibitor bort.

### 3.3. Rut enhances the inhibitory effect of bortezomib through synergism in WWP1

We observed that the binding affinity of the analog of commercial inhibitor Bort with WWP1 was lower than that of the single docking affinity of Car and Rut (Table 1). To explore potential enhancements in binding affinity and interactions for Bort with WWP1, we conducted MLSD with Bort and Rut. The binding free energy of MLSD was $-13.265 \pm0.25$ kcal/mol compared to the single docking results of Rut ($-10.26 \pm007$ kcal/mol) and Bort ($-7.64\pm0.156$). This enhanced binding affinity suggested a potential synergistic effect between the two ligands. We observed inter-ligand interactions, including hydrogen bonds and a pi-alkyl bond (Fig. 7d). From this, we hypothesized that both the ligands complement each other and exhibit synergy when docked together. The presence of these bonds is crucial in understanding the synergy, as hydrogen bonds contribute to the stability and specificity of the ligand interactions, while pi-alkyl bonds add to the overall stabilization of the complex through hydrophobic interactions (Li et al., 2023; Olszowy--Tomczyk, 2020). We hypothesized that when two hydrophobic molecules move together with inter-ligand interaction, their overall interaction with the complex is stabilized. These findings indicated that the presence of Rut may have enhanced the inhibitory effect of Bort, likely due to the stabilizing interactions between the ligands within the protein complex. Therefore, we hypothesized that both the ligands complement each other and exhibit synergism when docked together (Saha et al., 2024).

From MD analysis we observed that the RMSD of MLSD was higher compared to the single docked complexes and apo protein (Fig. 8a). This indicated greater structural fluctuations in the system due to the simultaneous presence of two ligands competing for binding at the same site. These competing interactions might have induced conformational changes in the protein-ligand complex, which lead to increased RMSD values, suggesting that the complex might not have been as stable as the apo protein. Despite these structural fluctuations in MLSD, we observed that the RMSD remained stable from 60 to 100 ns, indicating that the system eventually reached equilibrium during this period. Moreover, the lower Rg of MLSD compared to the apo protein suggested that the overall compactness of the complex was enhanced (Fig. 8b). This was likely because the two ligands, despite their competition, acted as stabilizing agents by promoting a compact configuration of the protein structure. The inter-ligand interactions and the additional interactions between the protein and the ligands could have reduced the movement of distant regions of the protein, thereby lowering the Rg. These observations suggested that while the presence of two ligands caused fluctuations, it also contributed to the overall structural stability and compactness of the docked complex. The MD simulation videos further support the hypothesis of synergism. In the MLSD video, Rut and the analog of Bort were strongly bound together in the same active site, indicating that they complement each other and act synergistically to inhibit the protein [see Supplementary video file 8]. Additionally, the MD video of analog of Bort single docking also showed stability [see Supplementary video file 9]. The per-residue energy decomposition analysis indicated that Leu 244, Tyr 259, Thr 473, Pro 477, Leu 484, Met 485 were common between MLSD and single dockings of analog of Bort and Rut (Fig. 8c). Moreover, Pro 477, Leu 484, Met 485 exhibited higher binding free energy in the case of MLSD. These observations suggested that the inter-ligand interactions between Car and Rut induced conformational changes in the protein structure, facilitating a more optimal binding configuration for the ligands. The total free energy of binding from MMPBSA was −19.98 kcal/mol for Bort, −20.15 kcal/mol for Rut, and −43.13 kcal/mol for MLSD. The free energy of MLSD exceeded the sum of the individual dockings, indicating the formation of a highly stable and strong complex with WWP1 through synergistic interactions. This enhanced stability suggested that Car and Rut together formed stronger interactions with the protein, effectively covering the regulatory site and potentially increased the inhibitory effect. These results highlighted the potential for synergistic interactions between our ligands and analog of Bort, offering insights into therapeutic strategies for targeting WWP1 in cancer treatment (Saha et al., 2024).

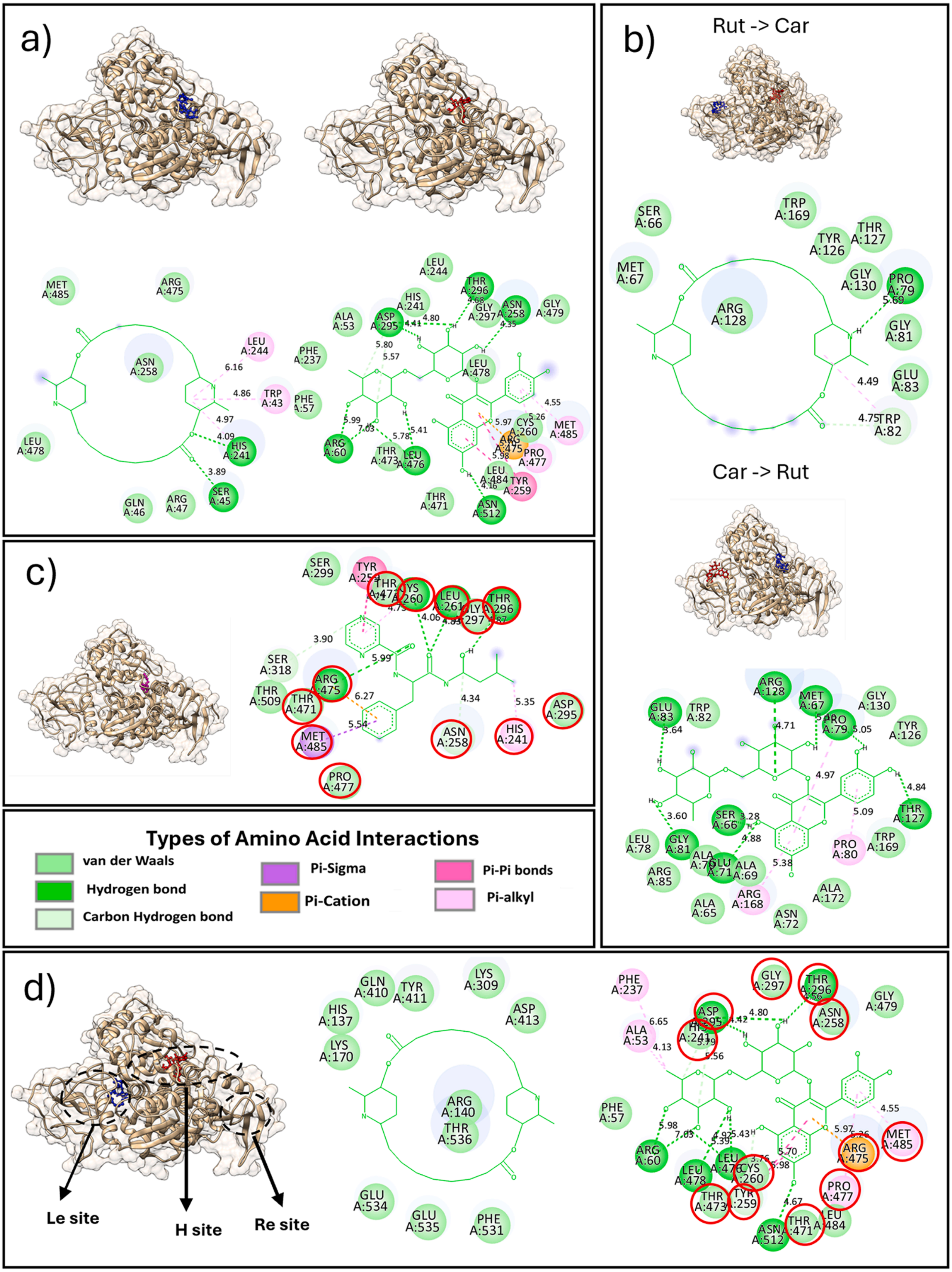

**Fig. 5.** Visualization of amino acid interactions and binding pocket of WWP1. (a) Single ligand docking of Carpaine (car) (blue) and Rutin (Rut) (red). They bind near the same protein pocket. (b) Sequential docking shows the presence of a second binding pocket. (c) Docking with bortezomib (Bort) analog. (d) MLSD and the regulatory sites of the protein have been high-lighted. The common amino acid interactions between Bort and Rut with WWP1 have been highlighted with red circles.

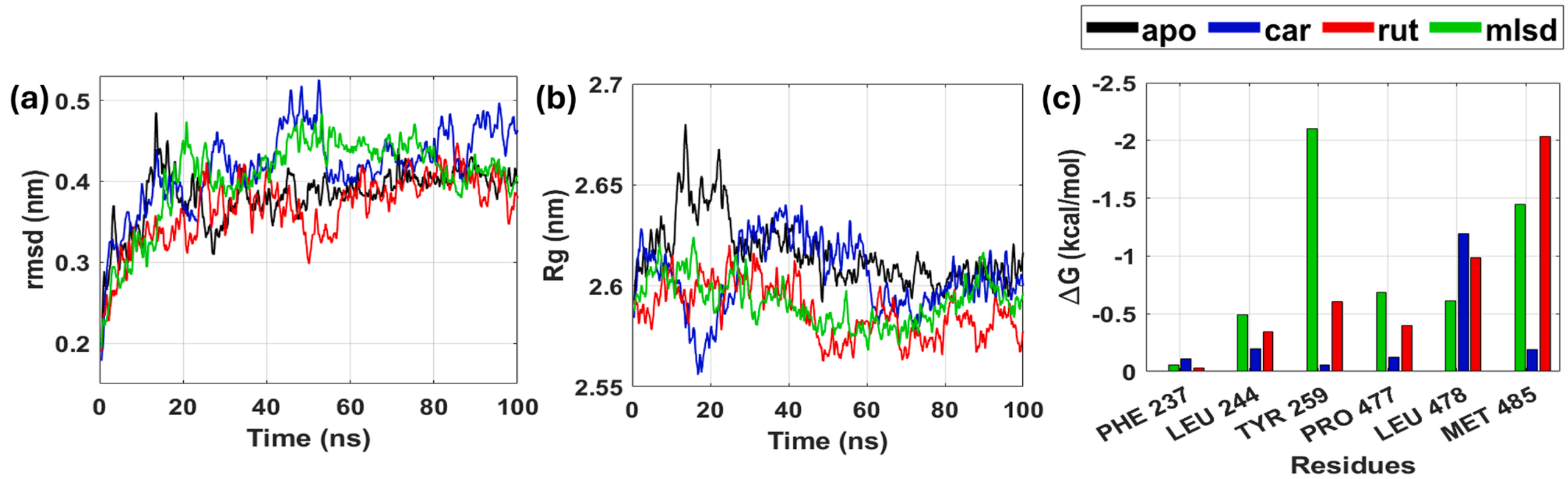

**Fig. 6.** (a) Root mean square deviation (RMSD) and (b) Radius of gyration (Rg) of all atoms computed for WWP1. (c) Free energy decomposition of residues for single dockings and MLSD with WWP1 allosteric effect.

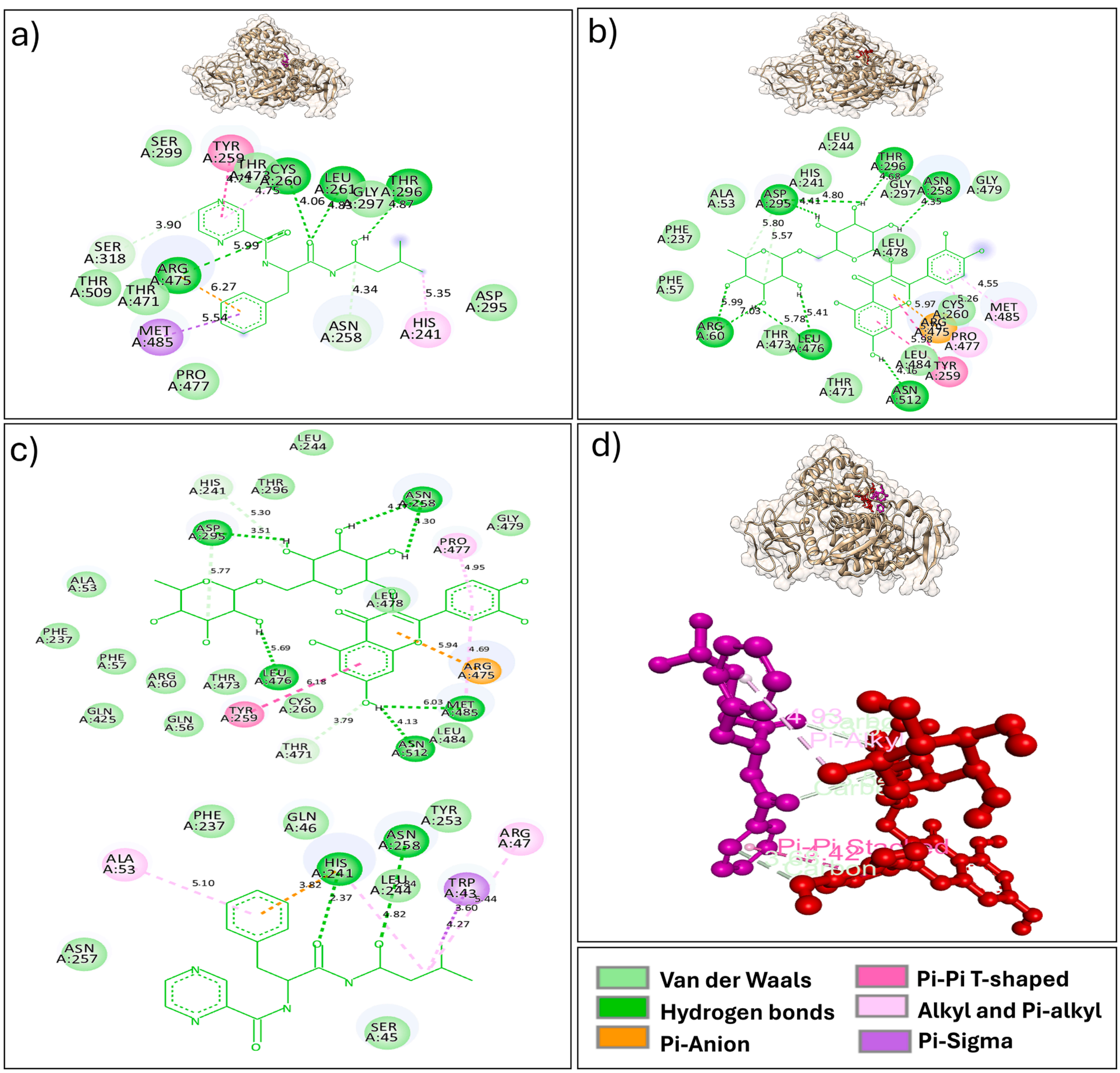

**Fig. 7.** Visualization of amino acid interactions and binding pocket of WWP1 (a) Single ligand docking with bortezomib (Bortezomib). analog (b) Single ligand docking with Rutin (Rut). (c) 2d visualization of Rut (top) and Bort analog (bottom). 3d visualization of MLSD with Bort analog (pink) and Rut (Red) (top), Interligand interactions between Bort analog and Rut (bottom).

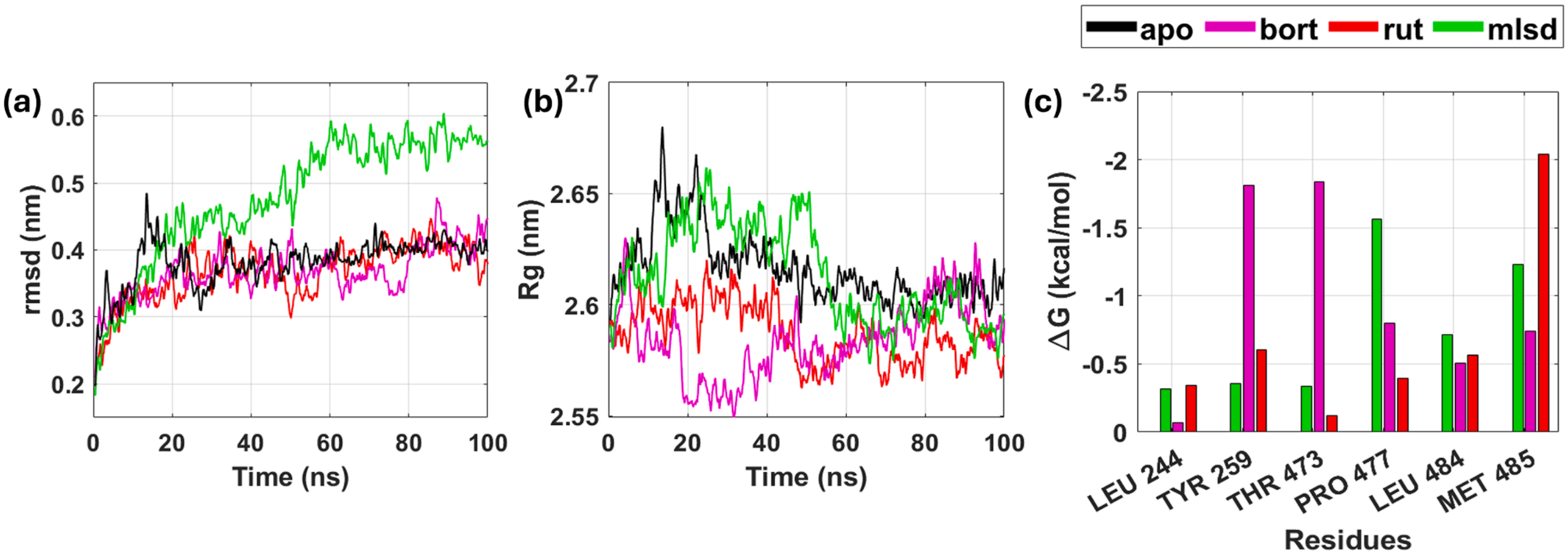

**Fig. 8.** (a) Root mean square deviation (RMSD) and (b) Radius of gyration (Rg) of all atoms computed for WWP1. (c) Free en-ergy decomposition of residue in WWP1 synergistic effect.

Several studies have focused on the synergistic effects of two ligands, by employing MLSD approach. One such study explored the synergism between baicalein and cubebin against the SARS-CoV-2 main protease (Mpro). Baicalein was identified as a primary active-site inhibitor, while cubebin acted as a synergistic allosteric partner. Molecular dynamics and free-energy calculations revealed that cubebin binding at two distant sites stabilized baicalein's interaction with Mpro. This study demonstrates that MLSD, combined with simulations and wavelet analysis, can uncover synergistic ligand crosstalk for novel combination-drug discovery (Li et al., 2023). Similarly, another study investigated the potential of 2-deoxy-d-glucose (2-DG) in combination with various drugs and phytochemicals for synergistic inhibition of SARS-CoV-2. The findings revealed that combinations with 2-DG exhibited stronger binding and greater stability than individual compounds, highlighting enhanced inhibitory potential. These results suggest that 2-DG-based combination therapies could be a promising approach for effective COVID-19 treatment (Gupta et al., 2022). In cancer research, a study by Aman et al. explored the synergistic potential of curcumin analogs AC01 and AC02 in combination with lapatinib (FMM), an ErbB4 inhibitor. Using MLSD and MD, the study compared the combinations with the reference FMM–5-fluorouracil (5-FU) pair. Unlike 5-FU, which binds outside the activation loop, AC01 and AC02, when combined with FMM, bound within the activation loop. While slightly weaker in binding than FMM–5-FU, the FMM–AC02 complex showed stronger interactions than FMM–AC01, suggesting that these analogs may enhance FMM's anticancer activity via synergistic binding (Aman et al., 2024). While these studies demonstrate the use of MLSD and MD simulations to establish synergistic effects for combination therapy, they often do not explore inter-ligand interactions. This is particularly important for phytochemicals, which remain less explored in combination therapies. The interaction between the two ligands is crucial as it can serve as leverage for combination therapy both ligands complementing each other and forming a stable complex. This increases the likelihood of both ligands binding together, thereby enhancing their combined therapeutic effect (Saha et al., 2024).

## 4. ADMET analysis

The ADMET results (Table 3) provided insights into the suitability of Car and Rut as drug molecules compared to commercial inhibitors. Car demonstrated high GI absorption like analog of Bort, while Rut exhibited low GI absorption as Ven. Plasma protein binding percentages for these ligands were comparable to commercial inhibitors, although their volume distribution was lower. Car had a moderate clearance rate (2.198 mL/min/kg), Rut showed a low clearance rate (1.349 mL/min/kg), Ven had 1.585 mL/min/kg, and an analog of Bort had the highest at 7.953 mL/min/kg. Lower clearance rates (Rut and Ven) suggest slower elimination and potential accumulation risks, while higher rates (analog of Bort) indicate faster elimination, possibly requiring more frequent dosing. Car's moderate clearance suggests a balance between elimination and systemic presence. Regarding toxicity, Car displayed respiratory and moderate skin toxicity, while Rut exhibited liver toxicity, which was also observed with commercial inhibitors. Ven showed moderate toxicity for the respiratory tract and carcinogenicity. Except for Rut and Ven, other compounds passed Lipinski's rule of five. Car and the analog of Bort demonstrated higher bioavailability (0.55), whereas Rut and Ven showed lower bioavailability (0.17). The predicted BCS class for Car and Rut was 4 and 3, respectively, which indicated challenges in formulation, but they can be improved through methods like nanonization (Khadka et al., 2014).

## 5. Limitations and future directions

Car and rut exhibited the potential to act as alternatives to commercial drugs for targeting cancer-related proteins. However, this study is based entirely on *in silico* methods, which, though useful for early-stage screening, cannot be used to completely replicate biological complexity. As such, *in vitro* and *in vivo* experiments are essential to confirm the additive, synergistic, and antagonistic effects of these phytochemicals and validate cytotoxicity, pharmacodynamics, and long-term safety. Although this substitution could theoretically affect docking accuracy, previous *in silico* studies have reported comparable docking results between bortezomib and its analogs, which further confirms the role of the analog as an accurate structural substituent. To address this, future research should employ software that supports such special atoms and include extensive *in vivo* and *in vitro* experiments to validate the findings. Additionally, nano formulations can be explored where phytochemicals are nanonized and incorporated into carriers such as nanoparticles or quantum dots, enhancing the efficiency of drug delivery. This approach allows for improved bioavailability, targeted delivery, and controlled release of the active compounds, optimizing their therapeutic potential.

**Table 3**
ADMET properties of Car and Rut compared with commercial inhibitors.

| Properties | Carpaine | Rutin | Venetoclax | Bortezomib analog |
|---|---|---|---|---|
| | Physiochemical properties | | | |
| Molecular weight (g/mol) | 478.380 | 610.52 | 868.44 | 356.42 |
| Number of heavy atoms | 34 | 43 | 61 | 26 |
| Number of aromatic heavy atoms | 0 | 16 | 27 | 12 |
| Fraction Csp3 | 0.93 | 0.44 | 0.38 | 0.37 |
| Number of rotatable bonds | 0 | 6 | 14 | 10 |
| Number of Hydrogen bond acceptors | 6 | 16 | 9 | 5 |
| Number of Hydrogen bond donors | 2 | 10 | 3 | 3 |
| Molar refractivity | 146.37 | 141.38 | 246.70 | 97.33 |
| Topological polar surface area (TPSA in Å$^2$) | 76.66 | 269.43 | 183.09 | 104.21 |
| | Absorption | | | |
| Water solubility | Moderately to Poorly soluble | Soluble | Insoluble to poorly soluble | Soluble |
| GI absorption | High | Low | Low | High |
| Skin permeability (cm/s) | −4.75 | −10.26 | −5.79 | −7.13 |
| P-glycoprotein substrate | No | Yes | Yes | Yes |
| Human intestinal absorption (HIA) | Low | Low | Low | Low |
| | Distribution | | | |
| Plasma Protein Binding | 71.516 % | 83.811 % | 101 % | 52.37 |
| Volume distribution | 0.728 | 0.754 | 1.041 | 1.387 |
| | Excretion | | | |
| Clearance (ml/min/kg) | 2.198 | 1.349 | 1.585 | 7.953 |
| Half-life (T1/2) | 0.126 | 0.524 | 0.012 | 0.525 |
| | Toxicity | | | |
| Skin sensitization | Moderately toxic | Nontoxic | Nontoxic | Nontoxic |
| Eye corrosion | Nontoxic | Nontoxic | Nontoxic | Nontoxic |
| Respiratory toxicity | Toxic | Nontoxic | Moderate | Nontoxic |
| Carcinogenicity | Non-toxic | Nontoxic | Moderate | Nontoxic |
| Drug-Induced Liver Injury (DILI) | Non-toxic | Toxic | Toxic | Toxic |
| Acute toxicity rule | 0 alerts | 0 alerts | 1 alert | 0 alerts |
| | Other parameters | | | |
| Lipinski rule | Yes | No | No | Yes |
| Bioavailability score | 0.55 | 0.17 | 0.17 | 0.55 |
| Synthetic accessibility | 6.82 | 6.52 | 6.05 | 3.42 |
| Lead likeness | No | No | No | No |
| | Predicted BCS classification | | | |
| Class | 4 | 3 | 2 | 3 |

## 6. Conclusion

Tumors often exhibit phenotypic variations that allow some cancer cells to survive and develop resistance, necessitating new drug development. This study explored the potential of two natural compounds, Car and Rut, derived from *Carica papaya* leaves, against proteins BCL-2 and WWP1. Molecular docking and dynamics simulations revealed that Car and Rut exert an additive effect on BCL-2 by binding to distinct regions within its binding pocket. Notably, their combined effect demonstrated improved binding affinity compared to individual ligands or the commercial inhibitor Venetoclax. For WWP1, Car binds near the H-site while Rut binds near the Le-site, showing an allosteric effect where one ligand enhances the other ligand's binding. Furthermore, the combination of Rut and analog of Bort exhibited a synergistic interaction with WWP1due to the presence of inter-ligand interactions that stabilized the complex. The free energy in MLSD exceeded the sum of the individual dockings, indicating the formation of a highly stable and strong complex with WWP1 through synergistic interactions. These findings highlight the potential of Car and Rut as effective phytochemicals for targeting BCL-2 and WWP1 in cancer therapy. Future research should validate these results through experimental assays compounds on receptor dynamics and interactions.

## Funding sources

This study was supported by the Dayananda Sagar University Seed Grant (Grant No. DSU/REG/2022-23/740).

## Glossary

Car,Carpaine; Rut, Rutin; Bort, Bortezomib; Ven,Venetoclax; MLSD, multi-ligand simultaneous docking; MD, molecular dynamics simulations; MMPBSA, Molecular Mechanics Pois-son-Boltzmann Surface Area; BCL-2, The B-cell lymphoma 2; WWP1, WW domain-containing E3 ubiquitin protein ligase 1; PDB, Protein Data Bank; SDF, Structure Data File; GI, gastro-testinal; BCS, Biopharmaceutical Classification System; FDA, food and drug administration; RMSD, Root Mean Square Deviation, TNBC, triple-negative breast cancer; CLL, chronic lymphocytic leukemia;SLL, small lymphocytic lymphoma; AML, acute myeloid leukemia; E1, ubiquitin-activating en-zyme; E2, ubiquitin-conjugating enzyme; E3, ubiquitin ligase; DOPE, Discrete Optimized Protein Energy; I3C, Indole-3-carbinol; 2DG, 2-deoxy-D-glucose; Mpro, main protease; FMM; lapatinib, 5-FU, 5-fluorouracil;

## CRediT authorship contribution statement

**Merla Sudha:** Writing – original draft, Formal analysis, Data curation. **Asmita Saha:** Writing – original draft, Formal analysis, Data curation. **Belaguppa Manjunath Ashwin Desai:** Writing – review & editing, Validation, Supervision, Funding acquisition, Formal analysis, Conceptualization. **Anil Ranu Mhashal:** Writing – review & editing, Validation, Formal analysis. **Pronama Biswas:** Writing – review & editing, Validation, Supervision, Funding acquisition, Formal analysis, Conceptualization.

## Declaration of competing interest

The authors declare that they have no known competing financial interests or personal relationships that could have appeared to influence the work reported in this paper.

## Acknowledgment

The authors thank AIC-DSU for supporting them with the server for conducting their study.

## Supplementary materials

Supplementary material associated with this article can be found, in the online version, at doi:10.1016/j.phyplu.2025.100829.